\documentstyle[11pt,epsf]{article}

\large
\newcommand{\be}{\begin{eqnarray}}
\newcommand{\ee}{\end{eqnarray}}
\newcommand\del{\partial}

\newcommand\Det{{\rm det}}
\newcommand\Tr{{\rm Tr\,}}
\newcommand\Real{{\rm Re}}
\newcommand\Imag{{\rm Im}}
\newcommand\str{{\rm Str \,}}
\newcommand\sdet{{\rm sdet}}
\newcommand{\ba}{\begin{array}}
\newcommand{\ea}{\end{array}}
\newcommand{\diag}{{\rm diag}}
\newcommand{\bC}{{\bf C}}
\newcommand{\bx}{{\bf x}}
\newcommand{\bj}{{\bf j}}

\begin{document}
\setlength{\baselineskip}{21pt}
\pagestyle{empty}
\vfill
\eject
\begin{flushright}
SUNY-NTG-97/11
\end{flushright}

\vskip 2.0cm
\centerline{\Large \bf  Universality of Correlation Functions}
\centerline{\Large \bf  in Random Matrix Models of QCD}
\vskip 1.5cm
\centerline{A.D. Jackson$^{a}$, M.K. \c Sener$^{b}$, J.J.M.
Verbaarschot$^{b}$}
\vskip 0.2cm
\centerline{$^{a}$The Niels Bohr Institute, Blegdamsvej 17,
            DK-2100 Copenhagen {\O}, Denmark}
\vskip 0.1cm
\centerline{$^{b}$Department of Physics, SUNY, Stony Brook, New York 11794}
\vskip 2cm

\centerline{\bf Abstract}

We demonstrate the universality of the spectral
correlation functions of a QCD
inspired random matrix model that consists of a random part having the
chiral structure of the QCD Dirac operator and a deterministic part which
describes a schematic temperature dependence. We calculate the correlation
functions analytically using the technique of Itzykson-Zuber integrals for
arbitrary complex super-matrices. An alternative exact calculation for
arbitrary matrix size is given for the special case of zero temperature, and we
reproduce the well-known Laguerre kernel. At
finite temperature, the microscopic limit of
the correlation functions are calculated in the saddle point approximation.
The main result of this paper is that the microscopic universality of
correlation functions is maintained even though unitary invariance is
broken by the addition of a deterministic matrix to the ensemble.

\vfill
\noindent

\eject
\pagestyle{plain}

\noindent
\section{Introduction}
\vskip 0.5cm

The study of level correlations in quantum physics has a long history.
It was first realized in nuclear physics that the spacing distribution
of compound nuclear resonances is given by random matrix theory
\cite{Porter}. Much later it was found numerically that the spectral
correlations of systems as simple as certain quantum
billiards \cite{berrytabor,bohigas} or polynomial potentials with
two degrees of freedom \cite{selig} can also be described by
random matrix theory.  The expectation that the spectral correlations
of the invariant random matrix ensembles are universal is most clearly
expressed by the Bohigas conjecture \cite{bohigas}: The spectral
correlations of a quantum system are given by random matrix theory if
the corresponding classical system is chaotic.  Although subsequent work
has provided considerable analytical insight regarding this conjecture
\cite{berry,bogomolny,andreev,kick,aleiner}, a proof has remained
elusive.  Meanwhile, the universality of random matrix correlations
has been tested in a wide variety of systems including the zeros of
the Riemann zeta function \cite{riemann}, sound modes in crystals
\cite{sound}, and the eigenvalues of the QCD Dirac operator \cite{HV}.

We address the question of universality within the context of random
matrix theory. An observable, typically a spectral correlation function,
will be called universal if it is stable against deformations of the
probability distribution.  Recently, this topic
has attracted a great deal of attention.  Two types of deformations have
been considered: deformations which maintain the invariance of the random
matrix ensembles \cite{Hack,arcmodels,loopeq} and those which violate that
invariance \cite{vwz,bz,hplusextern,zinnjustin}.
We will consider spectral correlation functions for an example of
the latter class.  In all cases which have been
studied, spectral correlations measured in units of the average level spacing
are found to be universal even for deformations that
change the spectral density on a macroscopic scale.

Specifically, we investigate the stability of the spectral correlations
of the chiral ensembles \cite{V2}. They are constructed to describe the
fluctuations of the Dirac eigenvalues in lattice QCD \cite{HV} and are
relevant to the theory of universal conductance fluctuations.
Because of an underlying $U_A(1)$ symmetry
(i.e. because the Dirac operator commutes with $\gamma_5$)
, the eigenvalues of chiral
matrices occur in pairs $\pm \lambda$. Therefore,
one can consider three types of universal behavior
each of which is given by invariant random matrix
ensembles:
(i) correlations in the bulk of the spectrum \cite{laguerre},
(ii) correlations near the edge
of the spectrum \cite{softedge},
and (iii) correlations near $\lambda = 0$.
In this case, the microscopic spectral density (i.e., the
spectral density near zero on the scale of a typical eigenvalue spacing)
is universal.  This has been illustrated for both invariant \cite{zee,ADMN}
and non-invariant deformations \cite{JSV1}.

The invariant random matrix ensembles and the chiral random matrix ensembles
are part of a larger classification scheme: Altland and Zirnbauer have
shown that there is a one to one correspondence between
random matrix ensembles and symmetric spaces \cite{class}.

The microscopic spectral density is of immediate physical interest.
According to the Banks-Casher formula \cite{BC}, the spectral density
at zero is directly proportional
to the order parameter of the chiral phase
transition in QCD (i.e., the chiral condensate).  The microscopic spectral
density provides information regarding the approach to the thermodynamic
limit.  This has been demonstrated, for example, in connection with the
dependence of the chiral condensate on the valence quark mass \cite{vplb}.
Recently, it has been verified by direct lattice calculations that
the microscopic spectral density of lattice QCD is given by one of
the chiral ensembles \cite{WGSW}. Moreover, the microscopic spectral density
enter in sum-rules \cite{LS} for the inverse eigenvalues of the Dirac operator.

The model which we shall consider is the chiral unitary ensemble perturbed
by the lowest Matsubara frequencies, $\pm \pi T$. This model was
introduced in \cite{JV} as a model for the chiral phase transition.  Indeed,
the average spectral density for this model undergoes a transition from
one semicircle at zero temperature to two disjunct semicircles at high
temperature.  In \cite{JSV1} it was shown that the spectral density of this
model on the scale of individual average level spacings is independent of
the temperature.

The aim of the present paper is to show that, below to the critical
temperature, {\em all\/} spectral correlations measured in units of
the average eigenvalue spacing are independent of the temperature.
This will be achieved using the methods
of super-symmetric integration introduced by Guhr \cite{Guhr91}
and certain super-symmetric Itzykson-Zuber integrals \cite{IZ, Mehta,
GW, JSV2}. At zero temperature our results coincide with earlier work
\cite{V2,Nagao-Slevin, VZahed,ADMN}.

\medskip

The outline of the paper is as follows. In section 2, we introduce the
random matrix model and express the correlation functions in terms of
a partition function.  In section 3, we reduce this partition function
to an integral over a supermatrix, $\sigma$, of much smaller size.  The
choice of an explicit parametrization for $\sigma$ is essential for the
rest of the paper.  Thus, we will argue in section 4 that the correlation
functions can be obtained by deforming the non-compact parametrization
required for uniform convergence into a compact parametrization of $\sigma$.
This will be done through a detailed investigation of the one-point function.
In section 5, we show that all correlation functions follow from a two-point
kernel.  This will be evaluated at zero temperature in section 6,
and we will reproduce the well-known Laguerre kernel.  In section 7, we
evaluate the two-point kernel at finite temperature using a saddle point
approximation.  Our primary result will be that the correlation functions
are independent of temperature except for a trivial rescaling of their
arguments.

\section{The random matrix model
and the partition function for correlations}

The partition function of Euclidean QCD is given by
\begin{eqnarray}
Z_E = \langle \Det ( i \gamma\cdot D + i M) \rangle_{S_E}
\label{ZE}
\end{eqnarray}
where $\gamma \cdot D$
is the Dirac operator, $M$ is the mass matrix, and the
brackets denote the average over gauge field configurations with respect
to the Euclidean Yang-Mills action $S_E$.
Using a lattice regularization with
anti-periodic boundary conditions in the time direction, we can expand
the fermion fields as a sum over Matsubara frequencies to obtain
\begin{eqnarray}
\psi({\bf x}, \tau) = \sum_{k=1}^N  \sum_{l=-n+1}^n
 \phi_k({\bf x}) \exp\left( \pi i (2 l - 1) T \tau \right) \ .
\label{psisum}
\end{eqnarray}
Here, $T$ is the temperature (i.e., the inverse length of the time
axis), $N$ is to be identified with the volume of space, and $n$ denotes
the total number of Matsubara frequencies retained in the expansion.
The $\phi_k$ are properly normalized spinors.

Using this basis, we separate the time derivative in the Dirac operator,
$\gamma_0 \del_\tau$, from the remaining terms. In terms of the matrix elements
of the Dirac operator  the partition function for $N_f$ massless flavors
can then be written as
\begin{eqnarray}
Z_{\rm gauge} = \left< \Det^{N_f} \left[
\left(\begin{array}{cc} 0 &i C^{\dagger} \\i C & 0
\end{array} \right)
+
\left(\begin{array}{cc}
0  & i\Theta \otimes {\bf 1}_N
\\ i\Theta \otimes {\bf 1}_N
&  0 \end{array} \right)
   \right]
    \right>_{S_E} \ .
\label{Zgauge}
\end{eqnarray}
The identity matrix of size $N$ is denoted by ${\bf 1}_N$, and $\Theta$
contains the Matsubara frequencies $\Theta = {\rm diag}\left(
-(2n-1)\pi T, \ldots, -\pi T, \pi T , \ldots, (2n-1)\pi T \right)$.
In (\ref{Zgauge}), we have used a chiral representation of the Dirac
matrices. For fundamental fermions with three colors, the matrix
$C$ is a complex matrix of size $ 2nN \times  2nN $.  The
detailed form of $C$ depends on the particular gauge field configuration.

The corresponding random matrix model is obtained by
replacing the matrix elements of $C$ by independent random variables with
a Gaussian distribution.
Thus, instead of (\ref{Zgauge}), we study the properties of
\begin{eqnarray}
Z_{\rm RM} = N_D \int d[C] e^{-N \Sigma^2 \Tr C^{\dagger} C}
\Det^{N_f} \left[
\left(\begin{array}{cc} 0 &i C^{\dagger} \\ iC & 0
\end{array} \right)
+
\left(\begin{array}{cc}
0  & i\Theta \otimes {\bf 1}_N
\\ i\Theta \otimes {\bf 1}_N
&  0 \end{array} \right)
   \right ] \ ,
 \label{Zrmt}
\end{eqnarray}
where the measure $d[C]$ is the Haar measure defined by
$\prod_{m,n}  d (\Real \, C_{mn}) d (\Imag \, C_{mn} )$, and $N_D$ is a
normalization constant to be specified later.  We wish to stress that
this partition function is a {\em schematic\/} model of the QCD
partition function. For example, we have ignored all spatial dependence
of the matrix elements of the Dirac operator, and the critical exponents
of this model are necessarily those of mean field theory.  It is
our claim, however, that this partition function belongs to the same
universality class as the QCD partition function with respect to local
spectral fluctuations.

In this paper, we will study the spectral correlation functions of the
model given by (\ref{Zrmt}).  For simplicity, we will restrict our
attention to the case in which we retain only the lowest Matsubara
frequencies, $\pm \pi T$.  This enables us to replace $\Theta$
by $\pi T $ using a unitary transformation.  After redefining $N$, the matrix
$C$ can be taken to be a complex $N \times N$ matrix.

As was shown in \cite{JV,JSV1}, this model shows a second-order phase
transition at $\pi T\Sigma = 1$. Below this temperature, chiral symmetry
is broken spontaneously with the chiral condensate given by
$\Xi =\Sigma\sqrt{(1-\pi^2 T^2 \Sigma^2)}$.  At all temperatures,
the spectral density follows from the solution of a cubic equation
\cite{JV,Stephanov,JSV1}. In particular, we mention that,
according to the Banks-Casher formula \cite{BC}, the spectral density
at zero virtuality is simply related to the chiral condensate:
\be
 \Xi =\frac{\pi\rho(0)}{N}\ \ .
\label{BC}
\ee
In \cite{JSV1} it was shown that the microscopic spectral density,
defined as
\be
\rho_S(u) = \lim_{N\rightarrow \infty} \frac 1{N\Xi}\rho(\frac u{N\Xi})
\label{microscopic}
\ee
is independent of the temperature parameter in the random matrix model.
This result greatly adds to our understanding of the empirically observed
universality of the microscopic spectral density in lattice QCD
\cite{vplb,WGSW} and instanton liquid simulations \cite{V2}. Because
of these results and numerical results for higher-order correlation functions
in the bulk of lattice QCD Dirac spectra \cite{HV}, we expect that the
higher-order microscopic correlation functions are universal as well.

The spectral density of a matrix with eigenvalues $\lambda_k$ is given by
\be
\rho(x) = \sum_k \delta(x-\lambda_k) \ .
\ee
The $k$-level correlation functions are then defined as
\be
R_k(x_1, \ldots, x_k) = \langle \rho(x_1) \ldots \rho(x_k) \rangle \ ,
\ee
where $\langle \cdots \rangle$ denotes the ensemble average with respect to
(\ref{Zrmt}). However, in our approach is is more convenient to work
with correlation functions of the resolvent
\be
G(x) =
\left\langle \Tr \frac {-1} {x + i \epsilon - D} \right\rangle \ ,
\ee
where $D$ is the random matrix Dirac operator
\be
D = \left ( \begin{array}{cc} 0 & C + \pi T \\  C^\dagger + \pi T  & 0
\end{array} \right ),
\ee
and $\epsilon$ is a positive infinitesimal.  The spectral density is
then given by
\be
\rho(x) = \frac 1 \pi {\rm Im \,} G(x).
\ee
The corresponding correlation functions are defined by
\be
\hat R_k(x_1, \ldots, x_k) =
\left\langle \prod_{l=1}^k  \frac 1 {\pi}\Tr \frac {-1}
{x_l + i \epsilon_l - D}
\right\rangle \ .
\label{Rkdef}
\end{eqnarray}
The spectral correlation functions follow
by taking the imaginary part of each trace.
In the following, we will drop the $i \epsilon_l$ term, assuming that $x_l$ has
a positive infinitesimal imaginary part.

Using the identity
\be
\Tr A^{-1} = - \frac{1}{2} \frac{\del}{\del j}
\left. \left[ \det( A-j) / \det( A+j) \right]  \right|_{j=0}\ ,
\ee
one can express the correlation functions in terms of a generating function
\begin{eqnarray}
\hat R_k(x_1, \ldots, x_k) =
\frac{1}{(2 \pi)^k} \left. \prod_{l=1}^k
 \frac{\del}{\del j_l}  Z_k(j_1, \ldots, j_k) \right|_{j_l=0} \ ,
\label{RZj}
\end{eqnarray}
\begin{eqnarray}
Z_k(j_1, \ldots, j_k) = N_D
\int d[C] e^{-N \Sigma^2 \Tr C^{\dagger} C}
\Det^{N_f} D\, \prod_{l=1}^k
\frac{ \Det(D-x_l+j_l)}{ \Det(D-x_l-j_l)} \ .
\label{Zjdef}
\end{eqnarray}
The prefactor, $N_D$, in (\ref{Zjdef}) is to be chosen such that $Z_k(0)=1$.
In the rest of the paper, we will restrict our attention to the quenched
problem with $N_f = 0$.  This is due to the fact that the current method
does not allow us to find a solution for arbitrary $N_f$.  The reason
for this will be discussed at the end of section 5.

\section{ Reduction of the partition function }

The expression (\ref{Zjdef}) for the generating function of correlators
involves an integral over a matrix, $C$, with $2 N^2$ degrees of freedom.
When the deterministic part, $\Theta$, is absent,
we can exploit the unitary invariance
\be
C \rightarrow U C V^{-1},
\ee
with $U$ and $V$ unitary matrices, in order to rewrite the partition function
as an integral over the eigenvalues of $C$ only.  It can then be evaluated
easily using, for example, the orthogonal polynomial method
\cite{Mehtabook, Nagao-Slevin}.  The presence of $\Theta$ destroys unitary
invariance, and standard methods are unsuitable.   However, unitary
invariance is not essential in the supersymmetric formulation of random
matrix theory \cite{Efetov,VWZrep}. Below, we adapt an approach based on the
supersymmetric method which is described in \cite{Guhr91}. This
method exploits the determinantal structure of the correlation functions
and allows us to calculate all $k$-point correlation functions at the same
time.  Readers not familiar with the use of supersymmetry methods in random
matrix theory are referred to \cite{Efetov,VWZrep, Guhr91}.

In this section, we will express the partition function, (\ref{Zjdef}),
as an integral over a $4k \times 4k$ supermatrix in which $N$ appears only
as an overall factor in the action.  This is particularly useful for the
investigation of the thermodynamic limit (i.e., $ N \rightarrow \infty$).

The ensemble average in (\ref{Zjdef}) is performed by writing the
determinants as Gaussian integrals over commuting and anticommuting
variables \cite{VWZrep}.  Equivalently, the product of determinants and
inverse determinants in (\ref{Zjdef}) can be considered as a superdeterminant
and can be written as a Gaussian integral over a complex supervector.
In order to do this we introduce the following supermatrices
\begin{eqnarray}
\bC & = &\diag (C, \ldots, C ; C, \ldots, C  )
\  , \hfill
\nonumber \\
\bx & = &\diag (x_1, \ldots, x_k ; x_1, \ldots, x_k ) \ ,
 \hfill \nonumber \\
\bj & = &\diag (-j_1, \ldots, -j_k ; j_1, \ldots, j_k) \ . \hfill
\label{Cxj}
\end{eqnarray}
Here, a semicolon separates the $k$ boson-boson blocks from the
$k$ fermion-fermion blocks.  Each block is of size $N \times N$.
Further, $x_p$ and $j_p$ are understood to be multiplied by the
$N \times N$ identity matrix.

In order to exploit the chiral structure of the problem, we introduce
a pair of complex supervectors, $\phi_A$ and $\phi_B$, each of size
$(k \, N ; k \, N )$.  The two numbers refer to the number of
commuting and anticommuting components.  Then, the product of
determinants in (\ref{Zjdef}) can be written as
\begin{eqnarray}
\prod_{l=1}^k
\frac { \Det(D-x_l+j_l)} { \Det(D-x_l-j_l)}
&=&
\sdet^{-1}
\left| \ba{cc}
 -\bx + \bj  & \bC^{\dagger} + \pi T  \\
\bC + \pi T  & -\bx + \bj
\end{array} \right|  \nonumber \hfill \\
&=&
\int d[\phi_A] d[\phi_B]
\exp \left [-i\left (\begin{array}{c} \phi_A^\dagger \\ \phi_B^\dagger
\end{array} \right)
\left( \ba{cc}
 -\bx + \bj  & \bC^{\dagger} + \pi T  \\
\bC + \pi T  & -\bx + \bj
\end{array} \right )\left (\begin{array}{c} \phi_A \\ \phi_B
\end{array} \right)\right ]
\ ,\nonumber\\
\label{Zphi}
\end{eqnarray}
where we have taken $\Theta = \pi T $.  In (\ref{Zphi}), the blocks
inside the superdeterminant refer to the chiral structure, i.e., each
block is itself a supermatrix.  The measure on the right hand side of
(\ref{Zphi}) is
\begin{eqnarray}
d[\phi] =
\prod_{p=1}^k
\left[ \prod_{l=1}^N
\frac{ d(\phi_{pl}^0)  d(\phi_{pl}^0)^\ast }{2 \pi}\right]
\prod_{q=1}^k
\left[ \prod_{m=1}^N
\frac{  d(\phi_{qm}^1) d(\phi_{qm}^1)^\ast }{  i  } \right] \ .
\label{dphi}
\end{eqnarray}
Here, $\phi$ refers to either $\phi_A$ or $\phi_B$; $\phi_{pl}^0$
and $\phi_{qm}^1$ denote the commuting and anticommuting components of
$\phi$, respectively.  The ranges of the indices follow from the limits in
the products.  The integrals with respect to the anticommuting
variables are normalized according to the convention \cite{Berezin},
$\int d\chi \, \chi =1$.  The constants in (\ref{dphi}) are chosen so
that Gaussian integrals do not result in additional prefactors.  We
recall that $x_l$ is understood to have a positive imaginary part, which
is sufficient to ensure the convergence of (\ref{Zphi}).

The ensemble average can now be performed immediately by substituting
(\ref{Zphi}) into (\ref{Zjdef}) and completing the square in the
exponent.  Interchange of the orders of the $C$- and
$\phi$-integration is justified because the $C$-integral is uniformly
convergent in $\phi$.  In order to proceed, we rewrite the terms
appearing in the exponent of (\ref{Zphi}):
\begin{eqnarray}
\phi_A^{\dagger}\cdot\bC^{\dagger}\cdot\phi_B
   =
 \sum_{\epsilon = 0}^1\sum_{p=1}^k (-1)^\epsilon
  \Tr \, C^{\dagger} \cdot
\left ( \phi_{Bp}^\epsilon \otimes \phi_{Ap}^{\epsilon\dagger}
 \right)   \ ,
\label{Zphiexp}
\end{eqnarray}
and the complex conjugate equation for
$\phi_B^{\dagger}\cdot\bC\cdot\phi_A$.
Here, $\phi_p^0$ and $\phi_q^1$ are commuting and anticommuting vectors
of length $N$ which represent the components of $\phi_A$ and $\phi_B$.
The $C$-integral can now be performed using
\begin{eqnarray}
\int d[C] \exp\left[ - N\Sigma^2 \Tr C^\dagger C
 - i\, \Tr (C^\dagger  X + C Y) \right]
= N_C^{-1} \exp\left[ - \frac 1 {N\Sigma^2}
 \Tr (X Y) \right]\ ,
 \label{fromCtophi}
\end{eqnarray}
where $N_C^{-1} = \int d[C] \exp\left[ - N\Sigma^2 \Tr C^\dagger C
\right]$.  (For $N_f = 0$ we have $N_C = N_D$ in (\ref{Zjdef}).)
The matrices are arbitrary complex $N\times N$ matrices.
Thus, the generating function (\ref{Zjdef}) becomes
\begin{eqnarray}
Z_k(j_1, \ldots, j_k)  &=& \,
\int d[\phi_A] d[\phi_B] \;
\exp \left [ - \frac 1 {N\Sigma^2}
 \sum_{\epsilon\epsilon' =0}^1
 (-1)^{\epsilon'}
 \sum_{p p'=1}^k (\phi_{Ap}^{\epsilon\dagger}\cdot\phi_{Ap'}^{\epsilon'})
                 (\phi_{Bp'}^{\epsilon'\dagger}\cdot\phi_{Bp}^\epsilon)
 \right]  \nonumber  \\
&\times&
\exp\left [ i\sum_{\epsilon =0}^1
 (-1)^{\epsilon} (
 \sum_{ p=1}^k (\bx-\bj)_p (\phi_{Ap}^{\epsilon\dagger}\phi_{Ap}^\epsilon +
\phi_{Bp}^{\epsilon\dagger}\phi_{Bp}^\epsilon)
-  \pi T (\phi_{Ap}^{\epsilon\dagger}\phi_{Bp}^\epsilon
+ \phi_{Bp}^{\epsilon\dagger}\phi_{Ap}^\epsilon) )\right].\nonumber\\
\label{Zphidef}
\end{eqnarray}

As a result of the ensemble average, we obtain a fourth-order term
in the exponent which should be decoupled using a Hubbard-Stratonovitch
transformation.  In order to accomplish this, we rewrite the first
exponential in (\ref{Zphidef}) as $\exp \left( -  \str ( A B )  /
{N\Sigma^2}\right)$.  The $(k+k)\times(k+k)$
super-matrix $A$ is given by
\begin{eqnarray}
A^{\epsilon\epsilon'}_{lm} = (\phi^{\epsilon\dagger}_{Al} \cdot
\phi^{\epsilon'}_{Am} )
\ee
with a similar form for $B$.  Now, we can `undo' the transformation in
(\ref{fromCtophi}) by introducing a non-hermitean complex $(k+k)\times(k+k)$
super-matrix, $\sigma$.
\begin{eqnarray}
\exp\left( - \frac 1 {N\Sigma^2} \str ( A B) \right)
=
  \, \int d[\sigma] \,
\exp\left[ - N\Sigma^2
  \str\left(  \sigma^\dagger \cdot \sigma \right)
 - i\, \str\left( \sigma^\dagger \cdot A
      + \sigma \cdot B \right) \right] \ .
\label{fromphitosigma}
\end{eqnarray}
After writing $\str( \sigma^\dagger \cdot A) = \phi^\dagger_A
\cdot ( \sigma^\dagger \otimes {\bf 1}_N ) \cdot \phi_A$ and
$\str( \sigma \cdot A) = \phi^\dagger_B \cdot ( \sigma \otimes {\bf 1}_N )
\cdot \phi_B$,
and performing
a shift of integration variables given
by $ \sigma \rightarrow \sigma + \bx - \bj $,
we can
express the generating function for the correlation functions as
\begin{eqnarray}
Z_k(j_1, \ldots, j_k)  =  \,
\int d[\phi_A] d[\phi_B] \;  \int d[\sigma] \,
\exp \left\{ -  N\Sigma^2
  \str\left[
  (\sigma^\dagger + \bx - \bj)
     \cdot
  (\sigma + \bx - \bj) \right] \right\}
\nonumber \\
\times \exp
  \left\{
  - i\phi^\dagger_A \cdot ( \sigma^\dagger \otimes {\bf 1}_N )
   \cdot \phi_A
  - i
\phi^\dagger_B \cdot ( \sigma \otimes {\bf 1}_N )
   \cdot \phi_B
  - i \pi T ( \phi_A^{\dagger}\cdot\phi_B +
          \phi_B^{\dagger}\cdot\phi_A)\right\} \ .
\label{beforethechange}
\end{eqnarray}
Here we still use the definition (\ref{Cxj}) for
${\bf x}$ and ${\bf j}$, however from this point
on $x_p$ and $j_p$ in (\ref{Cxj})
are {\em not} meant to be multiplied
by the $N\times N$ identity matrix.

At this point, we would
like to change the order of the integrations in (\ref{beforethechange})
and perform the Gaussian integrals in $\phi$.  This would result in a
superdeterminant involving the matrix $\sigma$.  However, this is possible
only if the $\phi$-integral is uniformly convergent in $\sigma$, which is
not the case if $\sigma$ is an arbitrary complex matrix. However, it can
be shown in general \cite{zirnall} that it is possible to choose a certain
non-compact parametrization of the $\sigma$-variables that ensures uniform
convergence.  An explicit construction for one- and two-point correlation
functions has been given in \cite{MIT,JSV1}. The parametrization
in the case of the one-point function will be discussed in great detail in the
next section.

Postponing further discussion of these points, we arrive at the following
expression after a change in the order of the integrals in
(\ref{beforethechange})
\begin{eqnarray}
Z_k(\bj) = \,
\int d[\sigma]
\exp \left(
 -N\Sigma^2 \str(\sigma^\dagger + \bx - \bj)
          \cdot (\sigma + \bx - \bj)
\right)
\sdet^{-N}
\left|\begin{array}{cc}
\sigma^{\dagger}  & \pi T \\
\pi T     & \sigma
\end{array} \right|   \ .
\label{Zsigma}
\end{eqnarray}
This partition function, with $N$ appearing only as an overall parameter,
is amenable to a saddle point approximation.

Before we continue with the evaluation of this partition function,
we will show in the case of the  one-point function that the imaginary
part of the partition function can be obtained by replacing the non-compact
parametrization of (\ref{Zsigma}) by a compact parametrization.  We will
conjecture that all higher-order multipoint correlators $R_k$, as opposed
to the $\hat R_k$ defined in (\ref{Rkdef}), can also be obtained by using
a compact parametrization with the $\sigma$-variables parameterized according
to
\be
\sigma = U S V^{-1} \ .
\ee
This choice of integration domain allows us to utilize the Itzykson-Zuber-like
integrals developed earlier \cite{IZ,Mehta,GW,JSV2} for the integration over
the super-unitary matrices.  Unfortunately, we have not been able to
construct a rigorous proof of this statement.

\section{The one-point function}

In this section, we wish to make the point that the spectral density can be
obtained from a compact parametrization of the $\sigma$-matrix rather than
the non-compact parametrization which is required for uniform
convergence, and which justifies the interchange of $\phi$- and
$\sigma$-integrations.  Since fermionic integrals are always finite,
convergence problems arise only from the integrations over the boson-boson
block of the $\sigma$ matrix.

The uniform convergence of the $\phi$-integrations can be achieved
if we perform a Hubbard-Stratonovitch transformation using the identity
\be
e^{-a^2+b^2} = \frac {-1}{\pi i} \int_{-\infty}^\infty
d s \int_{-\infty}^\infty
\sigma d\sigma e^{-\sigma^2 -2ia\sigma \cosh s - 2 ib\sigma \sinh s} \ ,
\ee
where $a$ is real positive and $b^2 - a^2$
has a negative real part.  When expressed in terms of
the real and imaginary parts of $\sigma_{BB} = \sigma_1 + i \sigma_2$,
this identity corresponds to the non-compact parametrization introduced
in \cite{JSV1}.  This parametrization is given as
\begin{eqnarray}
\sigma_1 &=& (\sigma-i\epsilon) \cosh s/\Sigma \ ,
\nonumber\\
\sigma_2 &=& i(\sigma-i\epsilon) \sinh s/\Sigma \ ,
\label{parbb}
\end{eqnarray}
where both $\sigma$ and $s$ run over the real line.  Note that this
parametrization covers only half of the $(\sigma_1,\sigma_2)$ plane.
(The parameter $\sigma$ introduced here should not be confused with the
matrix $\sigma$ defined in the previous section.)  We wish to contrast this
parametrization with the compact parametrization (i.e., in polar
coordinates),
\begin{eqnarray}
\sigma_1 &=&  \sigma \cos \theta
/\Sigma \ ,\nonumber \\
\sigma_2 &=& \sigma \sin \theta/\Sigma \ .
\label{compact}
\end{eqnarray}
After performing the integration over the Grassmann variables in the
$\sigma$-matrix, the partition function can be written as
\begin{eqnarray}
Z(j) &=& \frac {N^2}{\pi^2}
\int_{-\infty}^{\infty} \sigma   d\sigma \int_{-
\infty}^{\infty} d s
\int_0^{\infty}\rho d\rho \int_0^{2\pi} d\varphi
F((\sigma-i\epsilon)^2,\rho^2,t^2)\nonumber\\
&\times&
\left ( \frac {\rho^2 +t^2}{t^2-(\sigma-i\epsilon)^2}
 \right )^N e^{-N[\sigma^2
+ \rho^2 +2(x-j)\Sigma(\sigma-i\epsilon)
\cosh s + 2ix\Sigma\rho \cos
\varphi
+\Sigma^2((x-j)^2 -x^2)]} \ ,
\label{lastreal}
\end{eqnarray}
where $t=\pi T \Sigma$, $\rho$, and $\varphi$ parameterize the
fermion-fermion block of the $\sigma$ matrix.  We have displayed
the $i\epsilon$ dependence of (\ref{lastreal}) explicitly.
The convergence of the $s$-integral is guaranteed by choosing
$x-j$ to be pure imaginary.  The definition of the function
$F((\sigma-i\epsilon)^2,\rho^2,t^2)$ and additional details can be found
in \cite{JSV1}.

We perform the $\sigma$-integral first by saddle-point integration.
The $\sigma$-integrand has poles at $\pm t +i\epsilon$, which allows
us to deform the integration contour into the lower complex half-plane.
The saddle-points in the $\sigma$ integration are given by $\bar\sigma=
\pm i \sqrt{1-t^2}$, and thus only the saddle point with the negative sign
can be reached by deforming the integration contour. The resulting integral
can be analytically continued to $x-j$ just above the real axis.

We now consider the imaginary part of the partition function.  The
$\varphi$ integral gives rise to a Bessel function, $J_0$.  Thus, the
only contribution to the imaginary part comes from the factor
\begin{eqnarray}
S =\int_{-\infty}^\infty d s \exp[-2N(x-j)\Sigma\bar\sigma
\cosh s ] \ .
\end{eqnarray}
For positive $x$, we change integration variables according to $s
\rightarrow s-\pi i/2$. Then, the integration path can be deformed into
the integration path I shown in Fig. 1, and $S$ is given by
\begin{eqnarray}
S =\int_{\rm I} d s \exp[2iN(x-j)\Sigma\bar\sigma
\sinh s ] \ .
\end{eqnarray}

\begin{figure}
\epsfysize=6cm
\centerline{\epsfbox{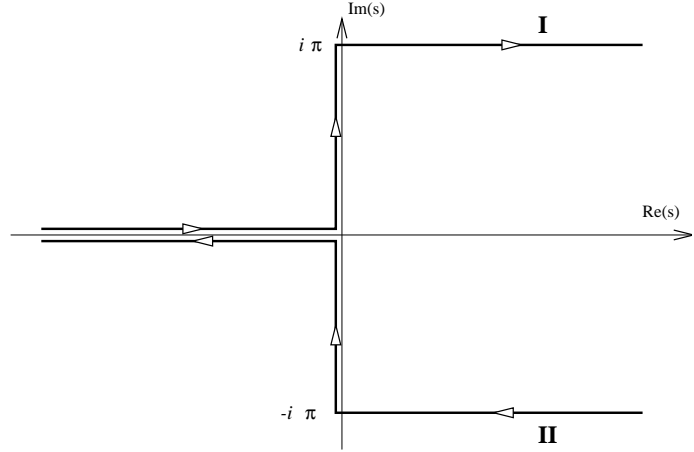} }
\caption[contour]{The integration contour for the
imaginary part of the generating function.}
\label{contour}
\end{figure}

The imaginary part of $S$ is given by half the sum (note the arrows)
of the contributions along the integration paths I and II in Fig.\,1.
The integral along the parts of the integration contour parallel to
the real axis cancel, and we are left with
\begin{eqnarray}
{\rm Im} \, S = \frac 12 (S - S^*) =
{\rm Im} \, \frac 12 \int_{-\pi i}^{\pi i} d s
\exp [2iN(x-j)\Sigma  \bar\sigma \sinh s  ] \ .
\end{eqnarray}
If we change integration variables, $s=i\theta$, we find
\begin{eqnarray}
{\rm Im} \, S = \int_{0}^{\pi} d\theta
\cos [2iN(x-j)\Sigma  \bar\sigma \sin\theta  ] \ .
\label{results}
\ee
In effect, this implies that the imaginary part of the partition function
can be obtained by replacing the $(\sigma, s)$ parametrization required
for uniform convergence by a compact parametrization such as (\ref{compact}).
With this parametrization, the partition function is given as
\be
Z^C(j) &=& \frac {N^2}{\pi^2}
\int_{0}^{\infty} \sigma   d\sigma \int_{0}^{2\pi} d\theta
\int_0^{\infty}\rho d\rho \int_0^{2\pi} d\varphi
F((\sigma-i\epsilon)^2,\rho^2,t^2)\nonumber\\
&\times&
\left ( \frac {\rho^2 +t^2}{t^2-(\sigma-i\epsilon)^2}
 \right )^N e^{-N(\sigma^2
+ \rho^2 +2(x-j)\Sigma(\sigma-i\epsilon) \cos \theta +
2ix\Sigma\rho \cos
\varphi
+\Sigma^2((x-j)^2 -x^2))} \ .
\label{zcompact}
\ee
In this form, the imaginary part of the partition function is
determined by the $i\epsilon$ in the singularities of the pre-exponential
factor.  Therefore, the imaginary part is even in $\sigma$.  This allows
us to extend the $\sigma$-integration from $-\infty$ to $\infty$ with
the introduction of a factor $1/2$.  Such an extension of the
$\sigma$-integration to the full real axis is required if we are to
perform this integral by saddle-point integration.  This integration
can be carried out in the same fashion as in the non-compact case and
yields the same result for ${\rm Im } \, Z^C(j)$ as obtained
by substituting the final result for ${\rm Im} \, S$ (\ref{results})
back into ${\rm Im } \, Z(j)$ defined in (\ref{lastreal}).

By contrast, the real part of the integral is odd in $\sigma$, and
it is not possible to extend the $\sigma$-integration over the entire
real axis.  Consequently, the $\sigma$-integration for the real part of $Z(j)$
cannot be performed by saddle-point integration if a compact parametrization
is adopted; only the imaginary part of $Z(j)$ can be obtained in this
fashion.  A similar phenomenon was observed in \cite{MIT} for both the
one- and two-point functions.

This leads us to the following conjecture: The spectral correlation
functions, which are given by the products of the imaginary parts of
the resolvents at different points, can be obtained by parameterizing the
$\sigma$-integration in (\ref{Zsigma}) as
\begin{eqnarray}
\sigma = U (S-i\epsilon) V^{-1} \ ,
\end{eqnarray}
where $U$ and $V$ are taken from the super-unitary group
as discussed in the next
section and where $S$ is a diagonal matrix.  This conjecture
is central to the remainder of this paper.

\section{Integration over soft modes using
Itzykson-Zuber integrals}

We now return to the evaluation of the generating function for correlations.
One possible method for the evaluation of (\ref{Zsigma}) would be to
integrate out anticommuting degrees of freedom first.  This is simple
for a small number of Grassmann variables when only a small number of terms
contribute to the integral. However, even for $k$ as small as 2 this
approach is impractical \cite{MIT}, and for larger values of $k$ it is
virtually impossible  to perform the integrals this way.

Rather, we adopt a technique developed in \cite{Guhr91}, where it
was applied to a study of the Gaussian unitary ensemble. This technique
consists essentially of separating the eigenvalue and angular coordinates
of $\sigma$ by diagonalization and then integrating over the angular
coordinates using a super-symmetric analogue of the Itzykson-Zuber integral.
The remaining bosonic
eigenvalue integrals can then be carried out either exactly (e.g.,
for the case of zero temperature considered in section 5) or in saddle point
approximation (e.g., for non-zero temperature as in section 6.)
The main virtue of Guhr's technique is
that it preserves the determinantal structure of the
correlation functions,
which makes it possible to obtain all correlation
functions at the same time.

We begin by reminding the reader of some
relevant properties of super-unitary matrices. As in the case of an arbitrary
complex matrix \cite{Hua},
an arbitrary super-matrix can be diagonalized by two
super-unitary matrices
\begin{eqnarray}
\sigma = U^{\dagger} S V \ ,
\label{diagsigma}
\end{eqnarray}
where $S = {\rm diag}( s^0_1, \dots, s^0_{k}; i s^1_1, \dots, i s^1_{k} )$
is a diagonal matrix which can be taken to have non-negative real entries
and $(U,V)$ parameterize the group $\left( U(k | k)\times U(k | k) \right)/
[ U(1) ]^{k+k}$.  (Here, $U(k|k)$ denotes the super-unitary group)
With this parametrization, the integration measure
can be written as
\footnote{
Note that our normalization of $d[\sigma]$ and
$d\mu(U,V)$ in  (\ref{IZint}) differs by a factor $2^{2k^2}$ from
ref. \cite{GW}. }
\begin{eqnarray}
d[\sigma] = d[S^2] d\mu (U,V) B^2(S^2) \ ,
\label{meas2}
\end{eqnarray}
where $d\mu (U,V)$ is the super-invariant Haar measure,
$d[S^2] = \prod_{\epsilon=0}^1
\prod_{l=1}^{k} d (s^\epsilon_l)^2 $.  The Jacobian of the
transformation is given by the square of the Berezinian \cite{GW}.
When the bosonic and fermionic blocks are of the
same size (i.e. for $N_f = 0$) the Berezinian can be
simplified to
\cite{Guhr91}
\begin{eqnarray}
B(S^2) = \det\left[ \frac 1 { (s^0_p)^2 - (i s^1_m)^2 }
 \right]_{p,m=1, \ldots, k}.
\label{berez0}
\end{eqnarray}

In terms of integration variables defined by the polar decomposition
(\ref{diagsigma}) (with an infinitesimal negative imaginary increment
included in the eigenvalues), the generating function (\ref{Zsigma})
can be rewritten as
\begin{eqnarray}
Z_k(\bj)   & = & \int  d[ S^2 ] \; B^2(S^2) \;
\sdet^{-N}
\left|\begin{array}{cc}
S  & \pi T \\
\pi T     & S
\end{array} \right|     \nonumber \\
& \times &
\int d\mu(U,V)  \exp\left(
-N\Sigma^2
\str (\sigma + \bx - \bj)^{\dagger} (\sigma + \bx - \bj)
\right) ,
\label{ZS}
\end{eqnarray}
where we have used the fact that
\begin{eqnarray}
\sdet
\left|\begin{array}{cc}
\sigma^{\dagger}  & \pi T \\
\pi T     & \sigma
\end{array} \right|
=
\sdet
\left|\begin{array}{cc}
S  & \pi T \\
\pi T     & S
\end{array} \right| \ .
\end{eqnarray}

The group integral in (\ref{ZS})
has precisely the form of an Itzykson-Zuber integral
over complex super-unitary  matrices.
For arbitrary complex super-matrices $\sigma= U^{-1} S V$ and
$\rho= U^{'-1} R V'$, one finds by an application of the heat kernel method
\cite{IZ, Mehta, GW, JSV2} that
\begin{eqnarray}
\int d\mu(U,V)e^{-\str[ (\sigma-
\rho)^{\dagger}(\sigma-\rho)]}
 =
\frac {1}{(k!)^2} \frac {
\det \gamma(s^0_p,r^0_q)
\det \gamma(s^1_m,r^1_n) }
{ B(S^2) B(R^2) } \ ,
\label{IZint}
\ee
where $(s^0_p; is^1_p)$ and $(r^0_p; ir^1_p)$ (with
 $p = 1, \ldots, k$) denote the eigenvalues of $\sigma$
and $\rho$, respectively. The quantity $\gamma(s,r)$ is defined as
\be
\medskip
\hfill\gamma(s,r) = \exp\left[-s^2-r^2 \right]I_0(2sr)\ .
\end{eqnarray}
For a derivation of this integral we refer the reader to
\cite{GW}. In this reference contributions
from Efetov-Wegner terms \cite{Rothstein} are discussed as well.

In (\ref{ZS}), the matrix $\rho$ is
diagonal with diagonal elements $R$ given by
\begin{eqnarray}
R = {\rm diag}(x_1+j_1, \ldots, x_{k}+j_{k} ;
  x_1-j_1, \ldots, x_{k}-j_{k}) \ ,
\label{Rdiag}
\end{eqnarray}
and the matrix $\sigma$ is as given in (\ref{diagsigma}).
It is easy to verify that
\begin{eqnarray}
\frac 1 { B(R^2)} =
\prod_{p=1}^k 4 x_p j_p  
 \times \left( 1 + {\sl O}(j)\right ) ,
\end{eqnarray}
which enables us to carry out the differentiations with respect to the source
terms. We observe that, as a result of (\ref{berez0}), the integrand of the
generating function (\ref{ZS}) factorizes into products
and determinants of
$k\times k$ matrices.  By renaming the integration variables $s^\epsilon_p$
it can be shown
that all $k!$ terms in the expansion of the determinants
of $\gamma(s^0_p,r^0_q) $ and $\gamma(s^1_m,r^1_n) $ are equal. The
only remaining determinant is the Berezinian (\ref{berez0}).
The products can then be absorbed into the
determinant, which leads to the following determinantal structure for
the correlation functions
\begin{eqnarray}
\hfill \hat R_k(x_1, \ldots, x_k)  & = &
 \det \left[ \Sigma\hat K_N(x_p\Sigma, x_q\Sigma) \right]_{p,q=1,\ldots, k}
\ ,
\label{Rkdet}
\ee
where the
dimensionless
kernel $\hat K_N(\xi, \eta)$ is given by
\be
\hfill \hat K_N(\xi, \eta)   &=&
-\frac{8N^2} \pi  \sqrt{\xi \eta}
\; \; \int_0^\infty d r \int_0^\infty d s
\frac {r s} {r^2 +s^2}
\left(  \frac {s^2 + t^2} {-r^2+t^2}
\right)^{N}   \hfill \nonumber \\
 &\times &  e^{-N(r^2+s^2 +(\xi^2-\eta^2)/2 )}
I_0\left(2N r \xi \right)
I_0\left(2N  i s \eta\right) \ .
 \hfill
\label{Rdet}
\end{eqnarray}
In this equation, we have introduced a dimensionless temperature
\be
t = \pi T \Sigma \ .
\ee
We have shown that all correlation functions follow from a single
two-point kernel. This is the main virtue of the application
of Guhr's supersymmetric method.

The spectral correlation functions $\hat R(x_1,\cdots,x_k)$ follow from
the discontinuities of $x_l$ across the real axis. The imaginary part
in (\ref{Rdet}) arises as a result of the $-i\epsilon$ term in the $s^0_p$.
One can easily convince oneself that the spectral
correlators preserve the determinantal structure (\ref{Rkdet}).
We thus find
\begin{eqnarray}
R_k(x_1, \ldots , x_k) & = &
\det\left[ \Sigma K_N(\Sigma x_p,\Sigma x_q)\right]_{p,q=1,\ldots,k}
 , \hfill
\ee
where
\be
\hfill K_N(\xi,\eta)   &=&
-\frac{8N^2}\pi \sqrt{\xi \eta}
\; \; \int_0^\infty d r \int_0^\infty d s
\frac {r s} {r^2 +s^2}
\left(   {s^2 + t^2}\right )^N  {\rm Im} \left (\frac 1{-(r-i\epsilon)^2+t^2}
\right)^N   \hfill \nonumber \\
 &\times &  e^{-N(r^2+s^2 +(\xi^2-\eta^2)/2 )}
I_0\left(2N  r \xi \right)
I_0\left(2N  i s \eta \right)  .
 \hfill
\label{Rkrealdef}
\end{eqnarray}
Both in (\ref{Rdet}) and (\ref{Rkrealdef}) the $(x,y)$-dependence has been
chosen so that the kernel is symmetric in $x$ and $y$ when $t=0$.

Both at $t = 0$ (section 6) and $t\ne 0$ (section 7)
the two integrals can be separated by the Feynman method,
\begin{eqnarray}
\frac 1 {r^2+s^2} =  N \int_0^\infty d\alpha
e^{- N\alpha (r^2+s^2)} \ .
\label{feynman}
\end{eqnarray}

Most of the results leading to (\ref{Rkdet}) can be generalized immediately
to an arbitrary number of flavors.  The main modifications are that the
fermionic blocks will be of size $k+N_f$ rather than $k$, and that the
supermatrices $\bx$ and $\bj$ in  (\ref{Cxj}) will have $N_f$ additional
zero blocks.  However, to the best of our knowledge, it is not possible to
write the Berezinian as a determinant as in (\ref{berez0}), and therefore
the determinantal structure of (\ref{Rkdet}) is lost.  This means that it
is no longer possible to express the correlation functions in terms of a
single kernel.

On the other hand, work on both Wigner-Dyson random matrix models
\cite{zinnjustin} and chiral random matrix models with arbitrary
unitary invariant potentials \cite{ADMN} using the orthogonal
polynomial method shows that the correlation functions still possess
a determinantal form. This suggests that the present approach might
be modified to arbitrary $N_f$ as well.

\section{Exact evaluation of correlation functions for
$T=0$}

In this section, we will show that kernel (\ref{Rkrealdef})
at $T=0$ reduces
to the usual Laguerre kernel obtained by means of the
orthogonal polynomial
method \cite{laguerre}.

After the introduction of new integration variables by
$u = r^2$ and $v=s^2$,
the kernel can be written as
\begin{eqnarray}
K_N(x,y) & = &
- \frac {2N^3} \pi \sqrt{x y}e^{-N(x^2-y^2)/2}
\int_0^\infty d\alpha {\cal I}_u(x) {\cal I}_v(y) \ , \hfill \nonumber \\
{\cal I}_u(x) & = & \int_0^\infty du
\,  {\rm Im} \left (\frac 1 {-u+i\epsilon} \right )^N
 I_0\left(2Nx\sqrt{u}\right)
 e^{-N(1+\alpha)u} \ , \nonumber \\
{\cal I}_v(x) & = & \int_0^\infty dv
  v^N
 I_0\left(i2Ny\sqrt{v}\right)
 e^{-N(1+\alpha)v} \ .
\label{CNexact}
\end{eqnarray}

In order to evaluate ${\cal I}_u$, we use the identity
\be
{\rm  Im\,} (u-i\epsilon)^{-N} = \pi\frac { (-1)^{N-1}}{(N-1)!}
\frac{\del^{N-1}}{\del u^{N-1}} \delta(u)
\ee
in (\ref{CNexact}) and apply partial integrations in order to eliminate
the derivatives of the delta function.  As a result, we find
\begin{eqnarray}
{\cal I}_u (x) = \left. \frac{\pi(-1)^N}{(N-1)!}
\frac{\del^{N-1}}{\del u^{N-1}} \right |_{u=0}
\left[ I_0\left( 2 N x\sqrt{u} \right)
   e^{-N(1+\alpha)u}\right] \ ,
\end{eqnarray}
which precisely describes the derivatives of the generating function
for the Laguerre polynomials.  (See eq.\,(\ref{A1}).)  We thus obtain
\begin{eqnarray}
{\cal I}_u (x) = -\frac {\pi ( N (1+\alpha))^{N-1}}{(N-1)!}
   L_{N-1} \left( \frac{N x^2}{1+\alpha} \right) \ ,
\label{Iu}
\ee
where $L_N$ are the Laguerre polynomials.  The integral
${\cal_I}_v$ is
well-defined and can be found in the literature.  (See eq.\,(\ref{A2}).)
The result is
\be
{\cal I}_v (x) = N! \frac{e^{-  \frac{N y^2}{(1+\alpha)}} }
{ (N(1+\alpha))^{N+1}}
 L_N \left( \frac{Ny^2}{1+\alpha} \right) \ .
\label{Iv}
\end{eqnarray}
After making a change of variables to $z=(1+\alpha)^{-1}$, we find for
the two-point kernel that
\begin{eqnarray}
K_N(x,y) = 2 N^2 \sqrt{x y} e^{-N(x^2-y^2)/2}\int_0^1 dz
e^{-Ny^2z} L_{N-1}(Nx^2 z) L_N(N y^2 z) \ .
\label{CNLs}
\end{eqnarray}
To further simplify (\ref{CNLs}), we will need the identity
(see eq.\,(\ref{A5}))
\begin{eqnarray}
e^{-z\eta} L_n(z \eta) L_{n-1} (z \xi) =  \frac d {dz}
\frac{ e^{-z \eta}
 \left(L_{n-1}(z \xi) L_n(z \eta) -
     L_n (z \xi) L_{n-1} (z \eta)
   \right) } {\eta - \xi} \ .
\label{magicofLs}
\end{eqnarray}
The integrand is now a total derivative, and we
reproduce the well-known result \cite{laguerre}
\begin{eqnarray}
K_N(x,y) = 2N \frac {\sqrt{x y}} {x^2 - y^2}
e^{-N (x^2+y^2)/2}
\left[ L_{N-1}(N x^2) L_{N}(N y^2)
  - L_{N}(N x^2) L_{N-1}(N y^2) \right] \ .
\label{CNfinal}
\end{eqnarray}
This justifies our claim that the spectral correlation functions
can be obtained by using a compact parametrization for the $\sigma$-variables.

It can be shown from the asymptotic properties of the Laguerre polynomials
that spectral correlations in the bulk of the spectrum are given by the
Gaussian unitary ensemble. The result for the microscopic region,
$\tilde x = N  x \approx {\cal O}(1 )$, follows from the asymptotic form
of the Laguerre polynomials,
\be
\lim_{n \rightarrow \infty}  L_n(\frac x n) =
 J_0(2 \sqrt x) \ ,
\label{asym}
\ee
which can be used after rewriting (\ref{CNfinal}) with the aid of recursion
relations for the Laguerre polynomials.  As a result, we find the microscopic
kernel
\begin{eqnarray}
K_S(\tilde x, \tilde y) =\lim_{N\rightarrow \infty} \frac 1{2N}
K_N(\frac{\tilde x}{N},\frac{\tilde y}{N})
=  \frac{\sqrt{\tilde x \tilde y}} {\tilde x^2 - \tilde y^2}
(  \tilde x J_0(2 \tilde y) J_1 (2 \tilde x) -
 \tilde y J_0(2 \tilde x) J_1 (2  \tilde y)  ) \ ,
\label{CmicroT0}
\end{eqnarray}
which agrees with results obtained previously
\cite{Nagao-Slevin,VZahed,MIT}.

Finally, the microscopic spectral density is given by
\begin{eqnarray}
\rho_S(\tilde x) & = & \lim_{y\rightarrow x} K_S(\tilde x,\tilde y) \\
 & = & \tilde x (
 J_0^2(2 \tilde x) + J_1^2(2\tilde x) ) \ ,
\label{rhomicroT0}
\end{eqnarray}
which is also in complete agreement with previous results.

\section{Correlation functions at nonzero temperature}

In this section we evaluate the microscopic limit of the imaginary part
of the two-point kernel at nonzero temperature and show that, up to a
scale factor, it is in agreement with the zero temperature result in
the limit $N\rightarrow \infty$. To this end, we perform the $u$ and $v$
integrals by a saddle point approximation.  However, the saddle-point
approximation to (\ref{Rkrealdef}) suffers from the difficulty
that $\bar u = -1+t^2$ and $\bar v = 1-t^2$.  As a result, the
pre-exponential factor diverges at the saddle-point. The aim of the
transformations performed in the first part of this section is to eliminate
this factor.

We begin by separating the $r$ and $s$ integrals (\ref{Rdet}) according to
the Feynman method (\ref{feynman}).  The Feynman parameter, $\alpha$, is
replaced by the new integration variable $\beta = 1/\sqrt{(1+\alpha)}$.  After
rescaling $r \rightarrow \beta r$ and $s \rightarrow \beta s$, we find
\be
\hfill \hat K_N(x,y)   &=&
-\frac {16N^2}\pi \sqrt{\tilde x \tilde y} \int_0^1\beta d\beta
 \int_0^\infty rd r \int_0^\infty sd s
\left(  \frac {s^2 + t^2/\beta^2} {-(r-i\epsilon)^2+t^2/\beta^2}
\right)^{N}   \hfill \nonumber \\
 &\times &  e^{-N(r^2+s^2) + (\tilde x^2-\tilde y^2 )/2}
J_0\left(2 i \beta r\tilde x \right)
J_0\left(2  \beta s \tilde y\right)\, ,\nonumber \\
 \hfill
\label{Chat6}
\end{eqnarray}
where we have also written the modified Bessel functions in terms of ordinary
Bessel functions.  Next, we express the product of the Bessel functions as
a derivative of the microscopic kernel (\ref{CmicroT0}) according to the
following remarkable identity (see eq.\,(\ref{A7}))
\begin{eqnarray}
2\beta\sqrt{x y}J_0(2 \beta x) J_0(2 \beta y) = \frac d {d \beta} \beta
K_S(\beta x,\beta y)\ .
\label{magicofJs}
\ee
This identity can be derived from eq.\,(\ref{magicofLs}) using the asymptotic
limit (\ref{asym}) of the Laguerre polynomials.  After insertion of this
identity in (\ref{Chat6}) and partial integration with respect to
$\beta$, we find
\be
\hfill \hat K_N(x,y)   &=&
-\frac{8N^2} \pi
 \int_0^\infty  d r \int_0^\infty  d s \sqrt{ r s}
e^{-N(r^2+s^2) +(\tilde x^2-\tilde y^2 )/2} \nonumber \\
&\times& \left \{
\left(  \frac {s^2 + t^2} {-(r-i\epsilon)^2+t^2}
\right)^{N} K_S(i r\tilde x, s\tilde y) - \int_0^1 d\beta
K_S(i\beta  r\tilde x, \beta  s\tilde y)\frac d{d\beta}
\left(  \frac {s^2 + t^2/\beta^2} {-(r-i\epsilon)^2+t^2/\beta^2}
\right)^N \right \} \ .
\hfill \nonumber \\
\label{C1C2}
\end{eqnarray}
The second term in this equation can be simplified further.  We
differentiate with respect to $s$ and undo the change of integration
variables at the beginning of this section, i.e., $ r \rightarrow r /\beta$,
$s \rightarrow s / \beta$, and $\alpha = (1-\beta^2)/\beta^2$.  Finally,
we perform the integration with respect to $\alpha$ and obtain
cancellation of the factor $r^2+s^2$ which results from the differentiation.
We find that the
integrands of the two terms in (\ref{C1C2}) differ only by a factor
$-t^2/(s^2+t^2)(-r^2+t^2)$.  Thus, this equation can be rewritten as
\be
\hfill \hat K_N(x,y)   &=&
-\frac {8N^2}\pi
 \int_0^\infty d r \int_0^\infty d s \sqrt{r s}
e^{-N(r^2+s^2) +(\tilde x^2-\tilde y^2 )/2} \nonumber \\
&\times&
\left(  \frac {s^2 + t^2} {-(r-i\epsilon)^2+t^2}\right)^{N}
\left( 1- \frac{t^2}{(s^2+t^2)(-r^2+t^2)}\right )
K_S(i r\tilde x, s\tilde y) \ .
\hfill \nonumber \\
\label{C1C2new}
\end{eqnarray}

The saddle point evaluation of $\hat K$ should be performed separately
in the microscopic limit and in the bulk of the spectrum.  In the latter
case, the asymptotic forms of Bessel functions enter in the saddle point
equations.  However, this is not the case in the microscopic limit with
$\tilde x$ fixed in the thermodynamic limit.  The saddle-point approximation
in the microscopic limit is particularly simple.  At the saddle point, we find
\be
\bar r^2 = -1 + t^2 \ ,\nonumber \\
\bar s^2 = 1-t^2 \ ,
\ee
and both the second derivatives with respect to $r$ and $s$ are
equal to $4N(1-t^2)$.
Note that the $\bar r^2$ is outside the integration domain.  As discussed
in section 4, the imaginary part of the integrand of $K_N(x,y)$ is an even
function of $r$, which allows us to extend the integration range from
$-\infty$ to $\infty$ at the cost of a factor of $1/2$.  It is then clear
that the integration path can be deformed to reach the saddle point
$-i \sqrt{1-t^2}$.  We cannot extend the integration path for the
evaluation of the real part of $K_N(x,y)$, and it is not clear how the
integrals can be performed by a saddle-point method.

As a result, we find
\begin{eqnarray}
\lim_{N\rightarrow\infty} \frac 1{2N}
\hat K_N(\frac {\tilde x}{N},\frac {\tilde y}{N})  =
i \zeta K_S(\zeta \tilde x, \zeta \tilde y) \ .
\label{C1C2final}
\ee
For convenience, we have introduced the scaling factor $\zeta =
\sqrt{1 - t^2}$, which gives the temperature dependence
of the spectral density in the neighborhood of $\lambda = 0$.  We have
shown that, up to this rescaling factor, the kernel $K_N$ is independent
of the temperature.  This constitutes the central result of this paper.

\section{Conclusions}

In this paper, we have shown that {\em all\/} correlations of the
eigenvalues near zero, measured in units of the average spacing, are
independent of temperature deformations of the unitary chiral random
matrix model.  This result extends previous work on the microscopic
spectral density.  Together with other recent work on the universality
of correlation functions with respect to deformations that preserve
unitary invariance, this firmly establishes the universality
of the complete eigenvalue distribution in the neighborhood of $\lambda = 0$.

It is our conjecture that the correlations of lattice QCD Dirac eigenvalues
near zero virtuality are in the universality class of the chiral Gaussian
Unitary Ensemble (chGUE). This conjecture has been supported by lattice
simulations of the average microscopic spectral density via the valence
quark mass dependence of the chiral condensate. In view of the present
results, it would be interesting to study the correlations of lattice
QCD Dirac eigenvalues in the neighborhood of $\lambda= 0$. Our prediction
is that such correlations are given by the chGUE.

The present results were obtained by a generalization to the chGUE of
the supersymmetric method developed by Guhr for the Gaussian Unitary
Ensemble.  The strength of this method is that it preserves the determinantal
structure of the correlation functions. As usual in the supersymmetric
formulation of random matrix theory, this method also requires a proper
parametrization of the integration variables.  Following work by
Wegner and Efetov, it was believed that hyperbolic symmetry was an
essential ingredient for the parametrization of the integration
manifold.  The surprising feature of the present method is
that all {\em spectral\/} correlation functions can be obtained
from a compact integration manifold.  One reason might be that, because
of $U_A(1)$ symmetry, the resolvent satisfies the relation
\be
G(x+i\epsilon) = -G(x-i\epsilon),
\ee
and that all spectral correlation functions can therefore be obtained
from a generating function that does not involve infinitesimal increments
of opposite signs.  However, this does not explain why Guhr's method also
allows for a compact integration manifold in the case of the GUE.
Clearly, more work is needed to address this issue.

Our results are based on the choice of a compact integration manifold.
We have provided two important pieces of evidence supporting this choice.
First, a detailed analysis of the imaginary part of the generating function
of the one-point function shows that a non-compact integration domain can
be transformed into a compact one.  Second, a compact parametrization of
the integration manifold reproduces the exact correlation functions of the
chGUE. However, the ultimate justification of this change of integration
variables remains an open problem.  We hope to
address this question in future work.

\medskip

\section*{Acknowledgements}
The reported work was partially supported by the US DOE grant
DE-FG-88ER40388.  After completion of the major part of this work, we
learned that Thomas Guhr and Tilo Wettig have simultaneously and
independently considered the present problem \cite{GWnew}.  We
would like to thank them for the patience which made joint publication of
this paper possible.

\medskip

\section*{Appendix A: Some useful identities}

\renewcommand{\theequation}{A.\arabic{equation}}
\setcounter{equation}{0}

A generating function for the generalized Laguerre polynomials is given by
\be
 \sum_{n=0}^\infty \frac {z^n}{\Gamma(n+\alpha + 1)}
L_n^\alpha(x) = e^z \frac 1{(x z)^{\alpha/2}}J_\alpha(2\sqrt{x z}) \ .
\label{A1}
\ee
The Laguerre polynomials are defined by $L_n(x) = L^{\alpha = 0}_N(x)$.
A closely related integral is given by
\be
\int_0^\infty dx x^{n+\alpha/2} e^{-a x} J_\alpha(2 b\sqrt x) =\frac{
n!}{a^{n+\alpha+1}} e^{-\frac{b^2}a }
L_n^\alpha\left (\frac{b^2}a \right) \ .
\label{A2}
\ee

The Laguerre polynomials satisfy  the following remarkable identity
\be
n L_{n-1}(z \eta) L_{n}(z \xi) =
(n-1) L_{n-1}(z \eta) L_{n-2}(z \xi)
 + e^{z \eta} \frac d {dz}
\left[ z e^{-z \eta} L_{n-1}(z \eta) L_{n-1}(z \xi) \right] \ .
\label{A3}
\ee
By a recursive application of this relation
and the Christoffel-Darboux formula,
\be
\sum_{k=0}^{n-1} L_k(\xi) L_k(\eta) = \frac n {\eta - \xi}
\left[ L_{n-1}( \xi) L_n( \eta) -
     L_n (\xi) L_{n-1} ( \eta) \right] \ ,
\label{A4}
\ee
we find
\begin{eqnarray}
e^{-z\eta} L_n(z \eta) L_{n-1} (z \xi) =  \frac d {dz}
\frac{ e^{-z \eta}
 \left(L_{n-1}(z \xi) L_n(z \eta) -
     L_n (z \xi) L_{n-1} (z \eta)
   \right) } {\eta - \xi} \ .
\label{A5}
\end{eqnarray}
From the asymptotic from of the Laguerre polynomials,
\be
\lim_{n\rightarrow \infty}n^{-\alpha} L_n^\alpha(\frac xn) = x^{-\alpha/2}
J_\alpha(2\sqrt x) \ ,
\label{A6}
\ee
we obtain the following relation for Bessel functions
\begin{eqnarray}
2\beta\sqrt{x y}J_0(2 \beta x) J_0(2 \beta y) = \frac d {d \beta} \beta
K_S(\beta x,\beta y)\ ,
\label{A7}
\ee
where the Bessel kernel $K_S$ is defined as
\be
K_S(x,y)
=  \frac{\sqrt{x y}} {x^2 - y^2}
\left(  x J_0(2 y) J_1 (2 x) -  y J_0(2 x) J_1 (2  y)  \right) \ .
\label{A8}
\ee
This kernel can be obtained from the Laguerre kernel
(\ref{CNfinal}) with the help of the asymptotic result (\ref{A7}) after
rewriting the Laguerre polynomials in the same order
by means of the recursion relation
\be
(\alpha + n)L^\alpha_{n-1} = x L^{\alpha +1}_n - (x-n) L^\alpha_n \ .
\label{A9}
\ee


\begin{thebibliography}{9}

\bibitem{Porter}
C.E. Porter, `{\it Statistical theories of spectra: fluctuations}',
Academic Press, 1965; R. Haq, A. Pandey, and O. Bohigas,
Phys. Rev. Lett. {\bf 48} (1982) 1086.

\bibitem{berrytabor}M.V. Berry and M. Tabor, Proc. Roy. Soc. London
{\bf A356} (1977) 375.


\bibitem{bohigas}O. Bohigas, M. Giannoni, and C. Schmidt,
Phys. Rev. Lett. {\bf 52} (1984) 1; O. Bohigas and M. Giannoni,
Lecture notes in Physics {\bf 209} (1984) 1.

\bibitem{selig}T. Seligman, J. Verbaarschot, and M. Zirnbauer,
Phys. Rev. Lett. {\bf 53}, 215 (1984); T. Seligman and J. Verbaarschot,
Phys. Lett. {\bf 108A} (1985) 183.

\bibitem{berry}M. V. Berry, Proc. Roy. Soc. London
{\bf A400} (1985) 229.

\bibitem{bogomolny} E.B. Bogomolny and J.P. Keating,
Phys. Rev. Lett. {\bf 77} (1996) 1472.

\bibitem{andreev}  A.V. Andreev, O. Agam, B.D. Simons and
B.L. Altshuler, Nucl. Phys. {\bf B482} (1996) 536.

\bibitem{kick} A. Altland and M. Zirnbauer,
Phys. Rev. Lett. {\bf 77} (1996) 4536.

\bibitem{aleiner}I.L. Aleiner and A.I. Larkin, {\it Role of divergence
of classical trajectories in quantum chaos}, cond-mat/9610034.


\bibitem{riemann}
A. M. Odlyzko, Math. Comp. {\bf 48}, (1987) 273.


\bibitem{sound}
C. Ellegaard, T. Guhr, K. Lindemann, H.Q. Lorensen, J. Nyg{\aa}rd,
and M. Oxborrow, Phys. Rev. Lett. {\bf 75} (1995) 1546.

\bibitem{HV}M. Halasz and J. Verbaarschot, Phys. Rev. Lett.
{\bf 74} (1995) 3920; M. Halasz, T. Kalkreuter, and J. Verbaarschot,
hep-lat/9607042, Lattice 1996.


\bibitem{Hack} G. Hackenbroich and H. Weidenm\"uller,
 Phys. Rev. Lett. {\bf 74} (1995) 4118.

\bibitem{arcmodels}J. Ambjorn and G. Akemann, J. Phys. {A29} (1996) L555;
G. Akemann, Nucl. Phys. {\bf B482} (1996) 403; S. Higuchi, C.Itoi,
S.M. Nishigaki, and N. Sakai, hep-th/9612237;
N. Deo, {\it Orthogonal polynomials and exact
correlation functions for two cut random matrix models}, cond-mat/9703136.

\bibitem{loopeq}S. Jain, Mod. Phys. Lett. {\bf A11} (1996) 1201.


\bibitem{vwz}
J. Verbaarschot, H. Weidenm{\"u}ller, and M. Zirnbauer,
Ann. Phys. (N.Y.) {\bf153} (1984) 367.


\bibitem{bz}
J. Ambjorn, J. Jurkiewicz, and Y. Makeenko, Phys. Lett. {B251} (1990)
517; E. Br\'ezin and A. Zee, Nucl. Phys. {\bf B402} (1993) 613; Nucl. Phys.
{\bf B424} (1994) 435; J. D'Anna, E. Br\'ezin, and A. Zee, Nucl. Phys.
{\bf B443} (1995) 433; E. Br\'ezin, S. Hikami, and A. Zee, {\it Universal
correlations for deterministic plus random hamiltonians}, hep-th/9412230;
C. Beenakker, Nucl.Phys. {\bf B422} (1994) 515.


\bibitem{hplusextern}T. Guhr, Phys. Rev. Lett. {\bf 76} (1996), 2258;
T. Guhr, {\it Transitions toward quantum chaos: with supersymmetry
from Poisson to Gauss}, Ann. Phys. (N.Y.) (in press), cond-mat/9510052;
E. Br\'ezin and S. Hikami, {\it An extension of level spacing universality},
cond-mat/9702213;

\bibitem{zinnjustin}
P. Zinn-Justin, {\it Random Hermitean matrices in an
external field}, cond-mat/9703033.

\bibitem{V2}
E. V. Shuryak and J. J. M. Verbaarschot, Nucl. Phys. {\bf A560}
(1993) 306; J.J.M. Verbaarschot, Nucl. Phys. {\bf B426} (1994) 559;
J. Verbaarschot, Phys. Rev. Lett. {\bf 72} (1994) 2531;
Phys. Lett. {\bf B329} (1994) 351; Nucl. Phys. {\bf B427} (1994) 434.

\bibitem{laguerre}
D. Fox and P. Kahn, Phys. Rev. {\bf 134} (1964) B1152; (1965) 228;
T. Nagao and M. Wadati, J. Phys. Soc. Jap. {\bf 60} (1991) 3298,
{\bf 61} (1992) 78, 1910; C.A. Tracy and H. Widom, Comm. Math. Phys.
{\bf 161} (1994) 289.


\bibitem{softedge}E. Kanzieper and V. Freilikher, Phys. Rev. {\bf E55}
{1997} 3712.


\bibitem{zee}E. Br\'ezin, S. Hikami, and A. Zee, Nucl. Phys.
{\bf B464} (1996) 411.

\bibitem{ADMN}
S. Nishigaki, Phys. Lett. {B387} (1996) 139; G. Akemann, P. H. Damgaard,
U. Magnea, and S. Nishigaki, {\it Universality of random matrices in the
microscopic limit and the Dirac operator spectrum}, hep-th/9609174.


\bibitem{JSV1}
A.D. Jackson, M.K. \c Sener, and J.J.M. Verbaarschot, Nuc. Phys.
{\bf B479} [FS] (1996) 707.

\bibitem{class}F. Dyson, Comm. Math. Phys. {\bf 19} (1970) 235;
A. Altland, M. Zirnbauer, Phys. Rev. Lett. {\bf 76} (1996) 3420;


\bibitem{BC}
T. Banks and A. Casher, Nucl. Phys. {\bf B169} (1980) 103.



\bibitem{vplb}J. Verbaarschot, Phys. Lett. {\bf B368} (1996) 137.


\bibitem{WGSW}
T. Wettig, T. Guhr, A. Sch{\"a}fer, and H. A. Weidenm{\"u}ller,
{\it The chiral phase transition, random matrix models, and
lattice data}, hep-ph/9701387; S. Meyer, private communication.

\bibitem{LS}
H. Leutwyler and A. Smilga, Phys. Rev. {\bf D46} (1992) 5607.

\bibitem{JV}A.D. Jackson and J.J.M. Verbaarschot, Phys. Rev.
{\bf D53} (1996) 7223.



\bibitem{Guhr91}T. Guhr, J. Math. Phys. {\bf 32} (1991) 336.

\bibitem{Rothstein}
M. J. Rothstein, Transactions of the American Mathematical
Society {\bf 299}, 387 (1987).

\bibitem{IZ}
C. Itzykson and J.B. Zuber, J.\,Math.\,Phys.\ {\bf 21} (1980) 411.

\bibitem{Mehta}
M.L. Mehta, Comm.\,Math.\,Phys.\ {\bf 79} (1981) 327.


\bibitem{GW}
T. Guhr and T. Wettig, J. Math. Phys. {\bf 37} (1996) 6395.


\bibitem{JSV2}
A.D. Jackson, M.K. \c Sener, and J.J.M. Verbaarschot, Phys. Lett.
{\bf B387} (1996) 355.

\bibitem{Nagao-Slevin}
K. Slevin and T. Nagao, Phys. Rev. Lett. {\bf 70} (1993) 635;
T. Nagao and K. Slevin, J. Math. Phys. {\bf 34} (1993) 2075;
J. Math. Phys. {\bf 34} (1993) 2317.

\bibitem{VZahed} J. J. M. Verbaarschot and I. Zahed, Phys. Rev. Lett.
{\bf 70} (1993) 3852.



\bibitem{GW90}
T. Guhr and H. A. Weidenm{\"u}ller, Ann. Phys. {\bf 199} (1990) 412.

\bibitem{Mehtabook}
M. L. Mehta, {\it Random Matrices}, 2nd. ed., Academic Press,
Boston, 1991.

\bibitem{Efetov}K.B. Efetov, Adv. Phys. {\bf 32} (1983) 53.

\bibitem{VWZrep}
J. Verbaarschot, H. Weidenm{\"u}ller, and M. Zirnbauer, Phys. Rep.
{\bf 129} (1985) 367.


\bibitem{Berezin}
F. A. Berezin, {\it Introduction to Superanalysis}, Kluwer Academic
Publishers, 1987.

\bibitem{zirnall}
M. Zirnbauer, J. Math. Phys. {\bf 37} (1996) 4986.


\bibitem{MIT}
A.V. Andreev, B.D. Simons, and N. Taniguchi, Nucl. Phys {\bf B432 [FS]}
(1994) 487.

\bibitem{Hua}
L. Hua, {\em Harmonic Analysis of Functions of Several Complex Variables in
the Classical Domain}, trans.\ L. Ebner and A. Kor\'ani, American
Mathematical  Society, Providence, RI, 1963.

\bibitem{GWnew}
T. Guhr and T. Wettig, to be published.

\bibitem{Stephanov}
M. Stephanov, Phys. Lett. {\bf B275} (1996) 249.


\end{thebibliography}
\end{document}